\documentclass[sigconf,natbib=false]{acmart}

\AtBeginDocument{%
  \providecommand\BibTeX{{%
    \normalfont B\kern-0.5em{\scshape i\kern-0.25em b}\kern-0.8em\TeX}}}

\usepackage{amsmath}
\usepackage{amssymb}
\usepackage{array}
\usepackage{booktabs}
\usepackage{color}
\usepackage{comment}
\usepackage{float}
\usepackage{flushend}
\usepackage[toc, acronym]{glossaries} 
\usepackage{graphicx}
\usepackage{ifpdf}
\usepackage{keyval}
\usepackage{listings}
\usepackage{microtype}
\usepackage{moresize}
\usepackage{multirow}
\usepackage{paralist}
\usepackage{rotating}
\usepackage{setspace}
\usepackage{setspace}
\usepackage{soul}
\usepackage{subfig}
\usepackage{tabularx}
\usepackage{url}
\usepackage{paralist}
\usepackage{wrapfig}
\usepackage{xcolor}
\usepackage{xspace}
\usepackage{makecell}

\definecolor{listinggray}{gray}{0.95}
\definecolor{darkgray}{gray}{0.7}
\definecolor{commentgreen}{rgb}{0, 0.4, 0}
\definecolor{darkblue}{rgb}{0, 0, 0.4}
\definecolor{middleblue}{rgb}{0, 0, 0.7}
\definecolor{darkred}{rgb}{0.4, 0, 0}
\definecolor{brown}{rgb}{0.5, 0.5, 0}
\definecolor{dkgreen}{rgb}{0,0.5,0}
\definecolor{orange}{rgb}{1,.5,0}
\definecolor{dandelion}{cmyk}{0,0.29,0.84,0}
\definecolor{mauve}{rgb}{0.58,0,0.82}
\definecolor{mygreen}{rgb}{0,0.6,0}

\newif\ifdraft
\drafttrue

\ifdraft
 \newcommand{\jhanote}[1]{\textcolor{red}  {***SJ: #1}}
 \newcommand{\vbnote}[1]{ {\textcolor{blue} { ***VB: #1 }}}
 \newcommand{\note}[1]{ {\textcolor{blue} { ***Note: #1 }}}
 \newcommand{\mrsnote}[1]{\textcolor{olive} { ***MRS: #1 }}
 \newcommand{\jdnote}[1]{ {\textcolor{orange} { ***JD: #1 }}}
 \newcommand{\tjnote}[1]{ {\textcolor{green} { ***TJ: #1 }}}
 \newcommand{\mtnote}[1]{ {\textcolor{purple} { ***MT: #1 }}}
\else
 \newcommand{\jhanote}[1]{}
 \newcommand{\vbnote}[1]{}
 \newcommand{\note}[1]{}
 \newcommand{\mrsnote}[1]{}
 \newcommand{\jdnote}[1]{}
 \newcommand{\tjnote}[1]{}
 \newcommand{\mtnote}[1]{}
\fi

\newcommand{\entk}{EnTK\xspace}

\newcommand{\B}[1]{\textbf{#1}\xspace}

\ifpdf
\DeclareGraphicsExtensions{.pdf, .jpg, .tif}
\else
\DeclareGraphicsExtensions{.ps,  .eps, .jpg}
\fi

\let\origdescription\description
\renewenvironment{description}{
  \setlength{\leftmargini}{0em}
  \origdescription
  \setlength{\itemindent}{0em}
  \setlength{\labelsep}{\textwidth}
}
{\endlist}

\usepackage[style=ACM-Reference-Format,backend=bibtex,sorting=none]{biblatex}
\begin{document}

\title{Adaptive Ensemble Biomolecular Applications at Scale}

\author{Vivek Balasubramanian}
\affiliation{\institution{Department of ECE, Rutgers University}}

\author{Travis Jensen}
\affiliation{\institution{Department of ChBE, University of Colorado Boulder}}

\author{Matteo Turilli}
\affiliation{\institution{Department of ECE, Rutgers University}}

\author{Peter Kasson}
\affiliation{\institution{Biomedical Engineering, University of Virginia}}

\author{Michael Shirts}
\affiliation{\institution{Department of ChBE, University of Colorado Boulder}}

\author{Shantenu Jha}
\affiliation{\institution{ECE, Rutgers University and Brookhaven National Laboratory}}

\renewcommand{\shortauthors}{V. Balasubramanian et al.}

\setcopyright{acmlicensed}
\acmConference[ICPP '19]{ICPP '19: ACM International Conference on Parallel Processing}{Aug 5--8, 2019}{Kyoto, Japan}
\acmBooktitle{ICPP '19: ACM International Conference on Parallel Processing, Aug 5--8, 2019, Kyoto, Japan}
\acmYear{2019}
\copyrightyear{2019}

\begin{abstract}
Recent advances in both theory and methods have created opportunities to
simulate biomolecular processes more efficiently using adaptive ensemble
simulations. Ensemble-based simulations are used widely to compute a number of
individual simulation trajectories and analyze statistics across them.
Adaptive ensemble simulations offer a further level of sophistication and
flexibility by enabling high-level algorithms to control simulations based on
intermediate results. Novel high-level algorithms require sophisticated
approaches to utilize the intermediate data during runtime. Thus, there is a
need for scalable software systems to support adaptive ensemble-based
applications. We describe the operations in executing adaptive workflows,
classify different types of adaptations, and describe challenges in
implementing them in software tools. We enhance Ensemble Toolkit (EnTK) -- an
ensemble execution system -- to support the scalable execution of adaptive
workflows on HPC systems, and characterize the adaptation overhead in EnTK. We
implement two high-level adaptive ensemble algorithms -- expanded ensemble and
Markov state modeling, and execute upto $2^{12}$ ensemble members, on
thousands of cores on three distinct HPC platforms. We highlight scientific
advantages enabled by the novel capabilities of our approach. To the best of
our knowledge, this is the first attempt at describing and implementing
multiple adaptive ensemble workflows using a common conceptual and
implementation framework.

\end{abstract}

\maketitle

\section{Introduction}\label{sec:intro}
Current computational methods for solving scientific problems in biomolecular
science are at or near their scaling limits using traditional parallel
architectures~\cite{cheatham-cise15}. Computations using straightforward
molecular dynamics (MD) are inherently sequential processes, and
parallelization is limited to speeding up each individual, serialized, time
step. Consequently, \textit{ensemble-based} computational methods have been
developed to address these
gaps~\cite{comer2014multiple,Laio2002}
In these methods, multiple simulation tasks are executed concurrently, and
various physical or statistical principles are used to combine the tasks
together with longer time scale communication (seconds to hours) instead of
the microsecond to milliseconds required for standard tightly coupled parallel
processing.

Existing ensemble-based methods have been successful for addressing a number
of questions in biomolecular modeling~\cite{husic_2018}.
However, studying systems with multiple-timescale behavior extending out to
microseconds or milliseconds, or studying even shorter timescales on larger
physical systems will not only require tools that can support
100$\times$--1000$\times$ greater degrees of parallelism but also exploration
of \textit{adaptive} algorithms. In adaptive algorithms, the intermediate
results of simulations are used to alter following simulations. Adaptive
approaches can increase simulation efficiency by greater than a
thousand-fold~\cite{pande-jctc-2010} but require a more sophisticated software
infrastructure to encode, modularize, and execute complex interactions and
execution logic.

We define \textit{adaptivity} as the capability to change attributes that
influence execution performance or domain specific parameters, based on
runtime information. The logic to specify such changes can rely on a
simulation within an ensemble, an operation across an ensemble, or external
criteria, such as resource availability or experimental data. In most cases,
adaptive algorithms can be expressed at a high level, such that the adaptive
logic itself is independent of simulation details (i.e., external to MD
kernels like NAMD~\cite{phillips2005scalable} or 
GROMACS~\cite{abraham2015gromacs}). 
Adaptive operations that are expressed independent of the internal details of
tasks facilitate MD software package agnosticism and simpler expression of
different types of adaptivity and responses to adaptivity. This promotes easy
development of new methods while facilitating scalable system software and
its optimization and performance engineering~\cite{kasson2018adaptive}.

Adaptivity 
enables study of longer simulation durations to investigate larger physical
systems and to efficiently explore high dimensional energy surfaces in finer
detail. The execution trajectory of such applications cannot be fully determined
\textit{a priori}, but depends upon intermediate results. Adaptive algorithms 
``steer'' execution towards interesting phase space or parameters and thus 
improve sampling quality or sampling rate. To achieve scalability and
efficiency, such adaptivity cannot be performed via user intervention and hence
automation of the control logic and execution becomes critical.

To guide the design and implementation of capabilities to encode and execute
adaptive ensemble applications in a scalable and adaptive manner, we identify
two such applications from the biomolecular science domain as shown in
Figs.~\ref{fig:ee_worflow} and~\ref{fig:adap_msm_workflow}. Although each of
these biomolecular applications have distinct execution requirements, and
coordination and communication patterns among their ensemble members, they
are united by their need for adaptive execution of a large a number of tasks.


This paper makes five contributions: (i) identifies types of ensemble
adaptivity; (ii) enhances Ensemble Toolkit~\cite{balasubramanian2016icpp}
(\entk), an ensemble execution system,  with adaptive capabilities; (iii)
characterizes the cost of adaptive capabilities in \entk; (iv) implements two
high-level adaptive ensemble algorithms and executes upto $2^{12}$ ensemble
members, on thousands of cores on three distinct HPC platforms; and (v) discusses
scientific insight from these adaptive ensemble applications.

It is important to note that these contributions do not depend upon a specific
simulation package -- MD kernel, or otherwise. As a consequence, the
capabilities and results apply uniformly to adaptive ensemble applications
from multiple domains. To the best of our knowledge, this is the first
reported framework that supports the specification and implementation of
general-purpose adaptive ensemble applications.



Section~\ref{sec:related_work} describes existing and related approaches.
Section~\ref{sec:science_motiv} presents two science drivers that motivate the
need for large-scale adaptive ensemble biomolecular simulations. We discuss
different types and challenges in supporting adaptivity in
Section~\ref{sec:adaptivity}. In Section~\ref{sec:entk}, we describe the
design and implementation of \entk, and the enhancements made to address the
challenges of adaptivity. In Section~\ref{sec:exps}, we characterize the
overheads in \entk as a function of adaptivity types, validate the
implementation of the science drivers, and discuss scientific insight derived
from executing at scale.

\section{Related Work}\label{sec:related_work}
Adaptive ensemble applications span several science domains including, but not
limited to, climate science, seismology, astrophysics, and bio-molecular
science. For example, Ref.~\cite{coulibaly2005nonstationary} studies adaptive
selection and tuning of dynamic RNNs for hydrological forecasting;
Ref.~\cite{behrens2005amatos} presents adaptive modeling of oceanic and
atmospheric circulation; Ref.~\cite{casarotti2007adaptive} studies adaptive
assessment methods on an ensemble of bridges subjected to earthquake motion;
and Ref.~\cite{lan2001dynamic} discusses parallel adaptive mesh refinement
techniques for astrophysical and cosmological applications. In this paper, 
we focus on biomolecular applications, as examples, employing algorithms to 
simulate biophysical events.

Algorithms consisting of one or more MD simulations, provide quantitative and
qualitative information about the structure and stability of molecular
systems, and the interactions among them. Specialized computer architectures
enable single MD simulations at the millisecond scale~\cite{shaw2008anton} but
alternative approaches are motivated by the higher availability of
general-purpose machines and the need to investigate biological processes at
the scales from milliseconds to minutes. Importantly, although we discuss
mostly biological applications, there are many applications of MD in material
science, polymer science, and interface science~\cite{KF3, Napolitano17}.

Statistical estimation of thermodynamic, kinetic, and structural properties
of biomolecules requires multiple samples of biophysical events. Algorithms
with ensembles of MD simulations have been shown to be more efficient at
computing these samples than single, large and long-running MD 
simulations~\cite{comer2014multiple,Laio2002,maragliano2014comparison,
chodera2006long}. Adaptive ensemble algorithms use runtime data to guide the
progression of the ensemble, achieving up to a thousand-fold increase in
efficiency compared to non-adaptive alternatives~\cite{Hinrichs2007,Singhal2005}.

Several adaptive ensemble algorithms have been formulated. For example,
replica exchange~\cite{mitsutake2004replica} consists of ensembles of
simulations where each simulation operates with a unique value of a sampling
parameter, such as temperature, to facilitate escape from local minima. In
generalized ensemble simulation methods, different ensemble simulations employ
distinct exchange algorithms~\cite{okamoto2004generalized} or specify diverse
sampling parameters~\cite{babin2008adaptively} to explore free-energy surfaces
that are less accessible to non-adaptive methods. In
metadynamics~\cite{Barducci2011} and expanded ensemble~\cite{Chelli2012},
simulations traverse different states based on weights ``learned'' adaptively.
Markov State Model~\cite{chodera2006long} (MSM) approaches adaptively select
initial configurations for  simulations to reduce uncertainty of the resulting
model.

Current solutions to encode and execute  adaptive ensemble algorithms fall
into two categories: monolithic workflow systems that do not fully support
adaptive algorithms and MD software packages where the adaptivity is embedded
within the executing kernels. Several workflow
systems~\cite{mattoso2015dynamic}, including Kepler,
Taverna
and
Pegasus
support adaptation capabilities only as a
form of fault tolerance and not as a way to enable decision-logic for
changing the workflow at runtime.

Well known MD software packages such as Amber, GROMACS and NAMD
offer capabilities to execute adaptive ensemble algorithms. However, these
capabilities are tightly coupled to the MD package, preventing users from
easily adding new adaptive algorithms or reusing the existing ones across
packages.


Domain-specific workflow systems such as Copernicus~\cite{pronk2015molecular}
have also been developed to support Markov state modeling algorithms to study
kinetics of bio-molecules. Although Copernicus provides an interactive and
customized interface to domain scientists, it requires users to manage the
acquisition of resources, the deployment of the system and the configuration
of the execution environment. This hinders Copernicus uptake, often requiring
tailored guidance from its developers.

Encoding the adaptive ensemble algorithm, including its adaptation logic
within MD software packages or workflow systems locks the capabilities to
those individual tools. In contrast, the capability to encode the algorithm
and adaptation logic as an user application promises several benefits:
separation between algorithm specification and execution; flexible and quick
prototyping of alternative algorithms; and extensibility of algorithmic
solutions to multiple software packages, science problems and scientific
domains~\cite{mckinley2004composing,kasson2018adaptive}. To realize these promises,
we develop the abstractions and capabilities to encode adaptivity at the 
ensemble application level, and execute adaptive ensemble applications
at scale on high performance computing (HPC) systems.



\section{Science Drivers}\label{sec:science_motiv}
In this paper, we discuss two representative adaptive ensemble applications from 
the biophysical domain: Expanded Ensemble and Markov State Modeling. Prior to
discussing the implementation of these applications, we describe the 
underlying algorithms.

\subsection{Expanded Ensemble}\label{ssec:usecase_ee}

Metadynamics~\cite{Barducci2011} and expanded ensemble (EE)
dynamics~\cite{Chelli2012} are a class of adaptive ensemble biomolecular 
algorithms, where
individual simulations jump between simulation conditions. In EE dynamics, the 
simulation states take one of \(N\) discrete states of interest, whereas in 
metadynamics, the simulation states are described by one or more continuous
variables. 
In both algorithms, each simulation explores the states independently. 
Additional weights are required to force the simulations to visit desired
distributions in the simulation condition space, which usually involves sampling
in all the simulation conditions. These weights are learned adaptively using a
variety of methods~\cite{Chelli2012}.



Since the movement among state spaces is essentially diffusive, the larger
the simulation state spaces, the more time the sampling takes. ``Multiple
walker'' approaches can improve sampling performance by using more than one
simulation to explore the same state space~\cite{comer2014multiple}. Further,
the simulation condition range can be partitioned into individual simulations
as smaller partitions decrease diffusive behavior~\cite{Janosi2009}. The
``best'' partitions to spend time sampling may not be known until after
simulation. These partitions can be determined adaptively, based on runtime
information about partial simulation results.

\begin{figure}[ht]
    \centering
    \includegraphics[width=0.48\textwidth]{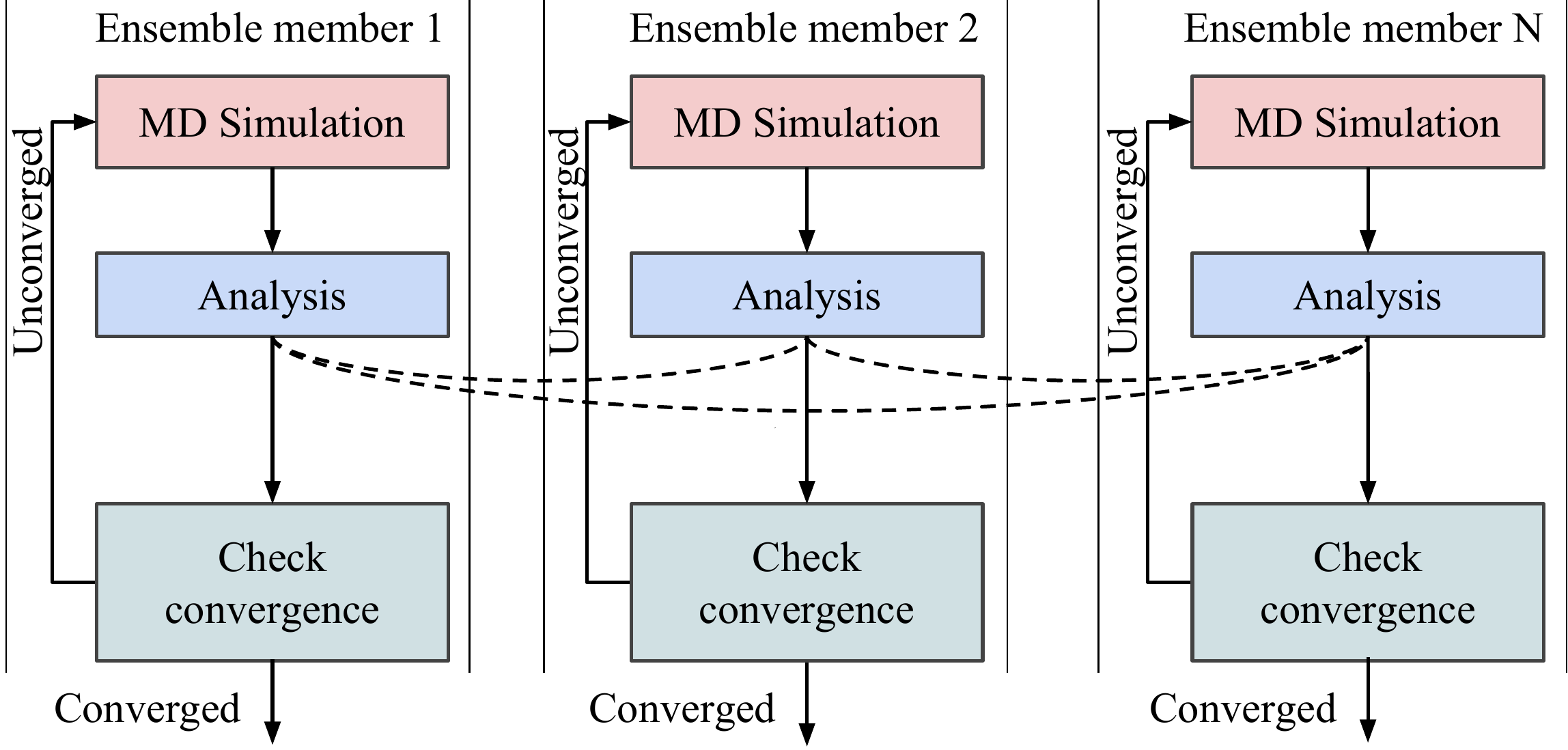}
    \Description{Three ensemble members, each with three stages. The first
    stage of each ensemble member contains a MD simulation task; the second
    stage contains an analysis task that uses the output of the MD simulation
    task of all ensembles; the third stage contains a task that checks
    whether convergence has been reached. If converged, the ensemble
    terminates, if not, a new MD simulation task is executed.}
    \caption{\footnotesize 
    Schematic of the expanded ensemble (EE) science
    driver. Two versions of EE are implemented: (1) local analysis where
    analysis only data local to its ensemble member; and (2) global analysis
    where analysis uses data from other ensemble members (represented by dashed
    lines)}\label{fig:ee_worflow}
\end{figure}

In this paper, we implement two versions of EE
consisting of concurrent, iterative ensemble members that analyze
data at regular intervals. In the first version, 
we analyze data local to each ensemble member; in the second version
we analyze data global to all the ensemble members by asynchronously
exchanging data among members. In our application, each ensemble member
consists of two types of task: simulation and analysis. The simulation tasks
generate MD trajectories while the analysis tasks use these trajectories
to generate simulation condition weights for the next iteration of simulation
in its own ensemble member. 
Every analysis task operates on the current snapshot of the total local or
global data. Note that in global analysis, EE uses any and all data available
and does not explicitly ``wait'' for data from other ensemble members.
Fig.~\ref{fig:ee_worflow} is a representation of these implementations.

\subsection{Markov State Modeling}\label{ssec:usecase_msm} 

Markov state modeling (MSM) is another important class of biomolecular simulation
algorithms for determining kinetics of molecular models. Using an assumption
of separation of time scales of molecular motion, the rates of first-order
kinetic processes are learned adaptively. In a MSM simulation, a large
ensemble of simulations, typically tens or hundreds of thousands, are run
from different starting points and similar configurations are clustered as
states. MSM building techniques include kinetic information but begin with a 
traditional clustering method (eg k-means or k-centers) using a structural 
metric. Configurations of no more than 2{\AA} to 3{\AA} RMSDs are typically 
clustered into the same ``micro-state''~\cite{pande2010everything}. 

The high degree of structural similarity implies a kinetic similarity,
allowing for subsequent kinetic clustering of microstates into larger
``macro-states''. The rates of transitions among these states are estimated
by observing which entire kinetic behavior can be inferred, even though
individual simulations perform no more than one state transition. However,
the choice of where new simulations are initiated to best refine the
definition of the states, improve the statistics of the rate constants, and
discover new simulation states requires a range of analyses of previous
simulation results, making the entire algorithm highly adaptive.

MSM provides a way to encode dynamic processes such as protein folding into a
set of metastable states and transitions among them. In computing MSM from
simulation trajectories, the metastable state definitions and the transition
probabilities have to be inferred. Refs.~\cite{Singhal2005,Hinrichs2007} show
that ``adaptive sampling'' can lead to more efficient MSM construction as
follows: provisional models are constructed using intermediate simulation
results, and these models are then used to direct the placement of further
simulation trajectories. Different from other approaches, in this paper we
encode this algorithm as an application where the adaptive code is
independent from the software packages used to perform the MD simulations and
MSM construction.

\begin{figure}[ht] 
    \centering
    \includegraphics[width=0.48\textwidth]{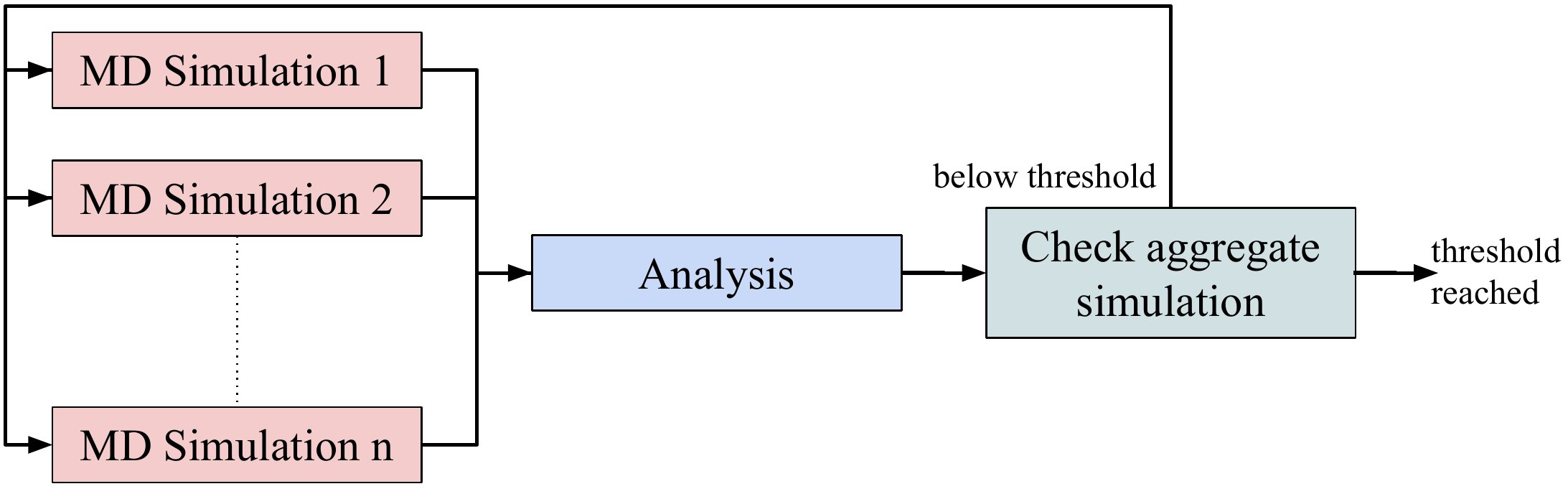}
    \Description{iterative pipeline with two stages: ensemble of simulations
    and MSM construction to determine optimal placement of future
    simulations. The first stage is repeated till sufficient amount of MD
    trajectory data is generated and analyzed. The analysis, i.e., second
    stage, operates over the cumulative trajectory data to generate a new set
    of simulation configurations, used in the next iteration of the
    simulations. The pipeline is iterated until convergence of the resulting
    Markov State Model.}
    \caption{\footnotesize 
    Schematic of the Markov State Model science
    driver.}\label{fig:adap_msm_workflow}
\end{figure}

Fig.~\ref{fig:adap_msm_workflow} offers a diagrammatic representation of the
adaptive ensemble MSM approach. The application consists of an iterative
pipeline with two stages: (i) ensemble of simulations and (ii) MSM
construction to determine optimal placement of future simulations. The first
stage generates sufficient amount of MD trajectory data for an analysis. The
analysis--i.e., the second stage--operates over the cumulative trajectory data
to adaptively generate a new set of simulation configurations, used in the
next iteration of the simulations. The pipeline is iterated until the
resulting MSM converges.

\section{Workflow Adaptivity}\label{sec:adaptivity}

Adaptive ensemble applications discussed in \S\ref{sec:science_motiv} involve
two computational layers: at the lower level each simulation or analysis is
performed via MD software package; at the higher level, an \textbf{algorithm}
codifies the coordination and communication among simulations and between 
simulations and analyses. Different adaptive ensemble applications and
adaptive algorithms might have varying coordination and communication patterns,
yet are amenable to common adaptations and similar types of adaptations.


We implement each simulation and analysis instance of these applications 
as a \textbf{task}, while representing 
the full set of task dependencies as task graph (TG) of a \textbf{workflow}. A
workflow may be fully specified a priori, or may be adapted, changing in
specification, during runtime. For the remainder of the paper, we refer to
alterations in the task graph as workflow adaptivity.



\subsection{Execution of Adaptive Workflows}\label{ssec:adap_exec}

\begin{figure}[ht]
  \centering
  \includegraphics[width=0.48\textwidth]{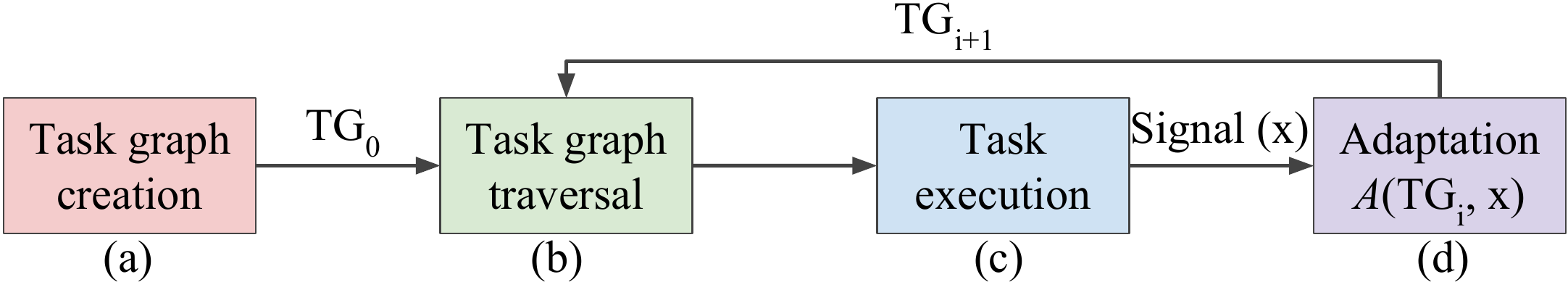}
  \Description{Four boxes, each representing an operation of the adaptive
  workflow: Task Graph Creation (TG_{0}); Task Graph Traversal; Task
  Execution (Signal (x)); and Adaptation (A(TG_{i},x)). }
  \caption{\footnotesize \textbf{Adaptivity Loop}: Sequence of operations in
  executing an adaptive workflow}\label{fig:adap_loop}
\end{figure}

Executing adaptive workflows at scale on HPC resources
presents several challenges~\cite{kasson2018adaptive}. 
%
%
Execution of adaptive workflows can be decomposed into four operations as
represented in Fig.~\ref{fig:adap_loop}: (a) creation of an initial TG,
encoding known tasks and dependencies; (b) traversal of the initial TG to
identify tasks ready for execution in accordance with their dependencies; (c)
execution of those tasks on the compute resource; and (d) notification of
completed tasks (control-flow) or generation of intermediate data (data-flow)
which invokes adaptations of the TG\@.

Operations (b)--(d) are repeated till the complete workflow is determined, and
all its tasks are executed. This sequence of operations is called an
Adaptivity Loop: in an adaptive scenario, the workflow ``learns'' its future
TG based on the execution of its current TG; in a pre-defined scenario, the
workflow's TG is fully specified and only operations (a)--(c) are necessary.

Encoding of adaptive workflows requires two sets of abstractions: one to
encode the workflow; and the other to encode the adaptation methods
(\textit{A}) that, upon receiving a signal \textit{x}, operate on the
workflow. The former abstractions are required for creating the TG, i.e.,
operation (a), while the latter are required to adapt the TG, i.e., operation
(d).

\subsection{Types of Adaptations}\label{ssec:adap_types}

Adaptivity Loop applies an adaptation method (Fig.~\ref{fig:adap_loop}d) to a 
TG\@. We represent a TG as $TG=[V,E]$, with the set V of vertices denoting
the tasks of the workflow and their properties (such as executable, required
resources, and required data), and the set E of directed edges denoting the
dependencies among tasks. For a workflow with $TG=[V,E]$, there exist four
parameters that may change during execution: (i) set of vertices; (ii) set of
edges; (iii) size of the vertex set; and (iv) size of the edge set. We
analyzed the $2^4$ permutations of these four parameters and identified
3 that are valid and unique. The remaining permutations represent conditions
that are either not possible to achieve or combinations of the 3 valid
permutations.

\paragraph{Task-count adaptation} We define a method $A_{tc}$ (operator) as a
task-count adaptation if, on receiving a signal \textit{x}, the method
performs the following adaptation (operation) on the TG (operand):

\begin{center}
$TG_{i+1}=A_{tc}(TG_i, x)$ 
$\implies size(V_i) \neq size(V_{i+1}) \land size(E_i) \neq size(E_{i+1})$ \\
where $TG_i = [V_i, E_i] \land TG_{i+1} = [V_{i+1}, E_{i+1}]$.
\end{center}

Task-count adaptation changes the number of TG's tasks, i.e., the adaptation
method operates on a $TG_i$ to produce a new $TG_{i+1}$ such that at least
one vertex and one edge is added or removed to/from $TG_i$.

\paragraph{Task-order adaptation} We define a method $A_{to}$ as a task-order
adaptation if, on a signal \textit{x}, the method performs the following
adaptation on the TG\@:

\begin{center}
$TG_{i+1}=A_{to}(TG_i, x)$ 
$\implies E_i \neq E_{i+1} \land V_i = V_{i+1}$ \\
where $TG_i = [V_i, E_i] \land TG_{i+1} = [V_{i+1}, E_{i+1}]$. 
\end{center}

Task-order adaptation changes the dependency order among tasks, i.e., the
adaptation method operates on a $TG_i$ to produce a new $TG_{i+1}$ such that
the vertices are unchanged but at least one of the edges between vertices is
different between $TG_i$ and $TG_{i+1}$.

\paragraph{Task-property adaptation} We define a method $A_{tp}$ as a
task-property adaptation if, on a signal \textit{x}, the method performs the
following adaptation on the TG\@:

\begin{center}
$TG_{i+1}=A_{tp}(TG_i, x)$ 
$\implies V_i \neq V_{i+1} \land size(V_i) = size(V_{i+1}) \land E_i = E_{i+1}$ \\
where $TG_i = [V_i, E_i] \land TG_{i+1} = [V_{i+1}, E_{i+1}]$. 
\end{center}

Task-property adaptation changes the properties of at least one task, i.e.,
the adaptation method operates on a $TG_{i}$ to produce a new $TG_{i+1}$ such
that the edges and the number of vertices are unchanged but the properties of
at least one vertex is different between $TG_i$ and $TG_{i+1}$.

We can represent the workflow of the two science drivers using the notations
presented. Expanded ensemble (EE) consists of $N$ ensemble members
executing independently for multiple iterations till convergence is reached
in any ensemble member. We represent one iteration of each ensemble members
as a task graph $TG$ and the convergence criteria with $x$. An adaptive EE workflow
can then be represented as:

\vspace{1mm}
\noindent\fbox{%
    \parbox{0.475\textwidth}{%
		\textit{parellel\_for} $i$ in $[1:N]$: 

			\hspace*{5mm} $while$ (condition on \textit{x}): 

			\hspace*{10mm} $TG_{i} =A_{tp}(A_{to}(A_{tc}(TG_{i})))$

		}%
}
\vspace{1mm}

Markov State Modeling (MSM) consists of one ensemble member which
iterates between simulation and analysis till sufficient trajectory data is
analyzed. We represent one iteration of the ensemble member as a task graph
$TG$ and its termination criteria as $x$. An adaptive MSM workflow can then
be represented as:

\vspace{1mm}
\noindent\fbox{%
    \parbox{0.475\textwidth}{%
	
	$while$ (condition on \textit{x}): 

		\hspace*{5mm} $TG = A_{to}(A_{tc}(TG))$ 
		}%
}

\subsection{Challenges in Encoding Adaptive Workflows}\label{ssec:challenges}

Supporting adaptive workflows poses three main challenges. The first
challenge is the expressibility of adaptive workflows as their encoding
requires APIs that enable the description of the initial state of the
workflow and the specification of how the workflow adapts on the base of
intermediate signals.
The second challenge is determining when and how to instantiate the
adaptation. Adaptation is described at the end of the execution of tasks
wherein a new TG is generated. Different strategies can be employed for the
instantiation of the adaptation~\cite{van2000dealing}. The third challenge is
the implementation of the adaptation of the TG at runtime. We divide this
challenge into three parts: (i) propagation of adapted TG to all components;
(ii) consistency of the state of the TG among different components; and (iii)
efficiency of adaptive operations.

\section{Ensemble Toolkit}\label{sec:entk}
EnTK is an ensemble execution system, implemented as a Python library, that
offers components to encode and execute ensemble workflows on HPC systems.
EnTK decouples the description of ensemble workflows from their execution by
separating three concerns: (i) specification of tasks and resource
requirements; (ii) resource selection and acquisition; and (iii) management
of task execution. EnTK sits between the user and the HPC system,
abstracting resource and execution management complexities from the user.

\begin{figure}[ht]
  \centering
  \includegraphics[width=0.48\textwidth]{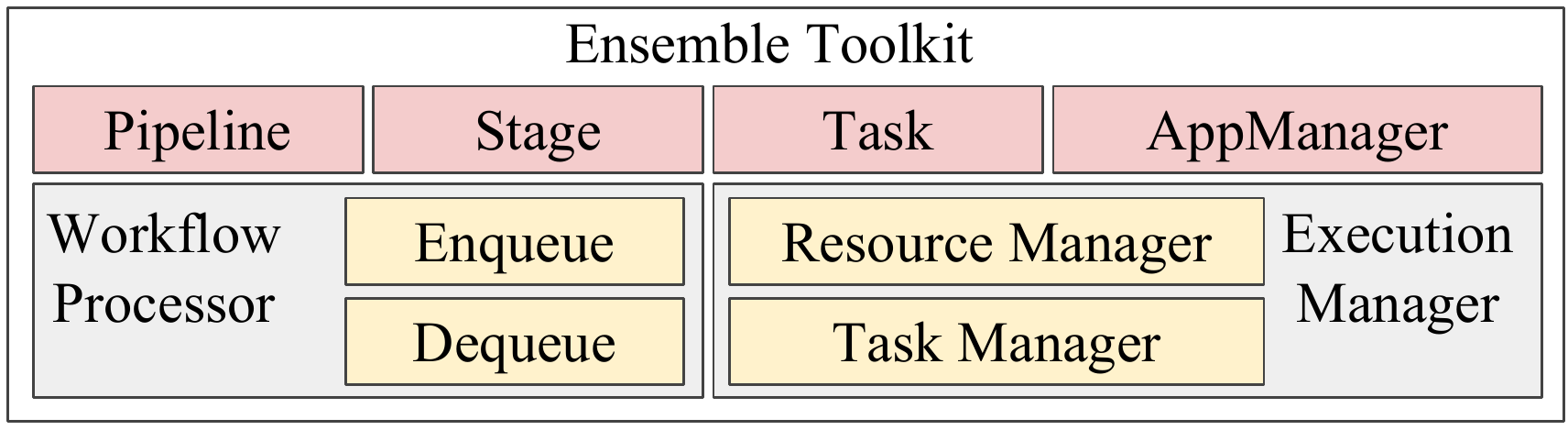}
  \Description{Two stack of boxes. The first stack has four red boxes, each
  representing an object of \entk API: Pipleine, Stage, Task, and AppManager.
  The second stack (below the first one) has two gray boxes, each
  representing an \entk component: WorkflowProcessor (with Enqueue and Dequeue
  modules); and ExecutionManager (with ResourceManager and TaskManager
  modules).}
  \caption{Schematic of EnTK representing its components and
  sub-components.}\label{fig:entk}
\end{figure}

EnTK is developed based on requirements elicited by use cases spanning
several scientific domains, including biomolecular, climate, and earth
sciences. The design, implementation and performance of EnTK is discussed in
detail in Ref.~\cite{power-of-many17}. We present a schematic representation
of EnTK in Fig.~\ref{fig:entk}, summarize its design and implementation, and
detail the enhancements made to EnTK to support the encoding and execution of
the three types of adaptation discussed in~\S\ref{ssec:adap_types}.

\subsection{Design}

EnTK exposes an API with three user-facing constructs: Pipeline, Stage, and
Task; and one component, AppManager. Pipeline, Stage, and Task are used to
encode the workflow in terms of concurrency and sequentiality of
tasks. We define the constructs as:

\begin{compactitem}
  \item \textbf{Task:} an abstraction of a computational process consisting
  of the specification of an executable, software environment, resource and
  data requirement.
  \item \textbf{Stage:} a set of tasks without mutual dependencies that,
  therefore, can be concurrently executed.
  \item \textbf{Pipeline:} a sequence of stages such that any stage
  \textit{i} can be executed only after stage \textit{i-1}.
\end{compactitem}

Ensemble workflows are described by the user as a set or sequence of
pipelines, where each pipeline is a list of stages, and each stage is a set
of tasks. A set of pipelines executes concurrently whereas a sequence
executes sequentially. All the stages of each pipeline execute sequentially,
and all the tasks of each stage execute concurrently. In this way, we
describe a workflow in terms of the concurrency and sequentiality of tasks,
without requiring the explicit specification of task dependencies.

AppManager is the core component of EnTK, serving two broad purposes: (i)
exposing an API to accept the encoded workflow and a specification of the
resource requirements from the user; and (ii) managing the execution of the
workflow on the specified resource via several components and a third-party
runtime system (RTS). AppManager abstracts complexities of resource
acquisition, task and data management, heterogeneity, and failure handling
from the user. All components and sub-components of EnTK communicate via a
dedicated messaging system that is set up by the AppManager.

AppManager instantiates a WorkflowProcessor, the component responsible for
maintaining the concurrent and sequential execution of tasks as described by
the pipelines and stages in the workflow. WorkflowProcessor consists of two
components, Enqueue and Dequeue, that are used to: enqueue sets of executable
tasks, i.e., tasks with all their dependencies satisfied; and dequeue
executed tasks, to and from dedicated queues.

AppManager also instantiates an ExecutionManager, the component responsible for
managing the resources and the execution of tasks on these resources.
ExecutionManager consists of two sub-components: ResourceManager and
TaskManager. Both sub-components interface with a RTS to manage the
allocation and deallocation of resources, and the execution of tasks,
received via dedicated queues, from the WorkflowProcessor.

EnTK manages failures of tasks, components, computing infrastructure (CI) and
RTS\@. Failed tasks can be resubmitted or ignored, depending on user
configuration. EnTK, by design, is resilient against
components failure as all state updates are transactional: failed components
can simply be re-instantiated. Both the CI and RTS are considered black boxes
and their partial failures are assumed to be handled locally. Upon full failure 
of the CI or RTS, EnTK assumes all the resources and the tasks undergoing 
execution are lost, restarts the RTS, and resumes execution from the
last successful pipeline, stage, and task.
\vspace{-1mm}
\subsection{Implementation}

EnTK is implemented in Python, uses the RabbitMQ message queuing
system~\cite{rmq} and the RADICAL-Pilot (RP)~\cite{merzky2018using} RTS\@. All
EnTK components are implemented as processes, and all subcomponents as
threads. AppManager is the master process spawning all the other processes.
Tasks, stages and pipelines are implemented as objects, copied among processes
and threads via queues and transactions. Process synchronization uses
message-passing via queues.

Using RabbitMQ offers several benefits: (i) producers and consumers are
unaware of topology, because they interact only with the server; (ii)
messages are stored in the server and can be recovered upon failure of EnTK
components; (iii) messages can be pushed and pulled asynchronously because
data can be buffered by the server upon production; and (iv) 
$\ge O(10^6)$ tasks are supported.

EnTK uses RP, a pilot system, as the RTS\@. Pilot systems enable the
submission of "pilot" jobs to the resource manager of an HPC system. The
defining capability is the decoupling of resource acquisition from task
execution. Pilot systems allow for queuing a single job via the batch system
and, once this job becomes active, it executes a system application that
enables the direct scheduling of tasks on the acquired resources, without
waiting in the batch system's queue. RP does not attempt to `game' the
resource manager of the HPC system: Once queued, the resources are managed
according to the system's policies. RP provides access to several HPC systems,
including XSEDE, ORNL, and NCSA resources, and can be configured to use other
HPC systems.


\subsection{Enhancements for Adaptive Execution}

In~\S\ref{ssec:challenges}, we described three challenges for supporting
adaptive workflows: (i) expressibility of adaptive workflows; (ii) when and
how to trigger adaptation; and (iii) implementation of adaptive
operations. \entk does not suppport these
adaptation requirements, nor can algorithms cannot be expressed. Therefore, we
engineered \entk with three new capabilities:
expressing an adaptation operation, executing the operation, and modifying a
TG at runtime.

Adaptations in ensemble workflows follow the Adaptivity Loop described
in~\S\ref{ssec:adap_exec}. Execution of one or more tasks is followed by some
signal \texttt{x} that triggers an adaptation operation. In \entk, this signal
is currently implemented as a control signal triggered at the end of a stage
or a pipeline. We added the capability to express this adaptation operation as
post-execution properties of stages and pipelines. In this way, when all the
tasks of a stage or all the stages of a pipeline have completed, the
adaptation operation can be invoked to evaluate based on the results of the
ongoing computation, whether a change in the TG is required. This is done
asynchronously without effecting any other executing tasks.

The adaptation operation is encoded as a Python property of the Stage and
Pipeline objects. The encoding requires the specification of three functions:
one function to evaluate a boolean condition over \texttt{x}, and two
functions to describe the adaptation, depending on the result of the boolean
evaluation.

\begin{figure}
\begin{lstlisting}[label=post_exec_code,caption={\footnotesize Post execution properties of
a Stage consisting of one Task. At the end of the Stage, 'function\_1'
(boolean condition) is evaluated to return a boolean value. Depending on
the value, 'function\_2' (true) or 'function\_3' (false) is invoked.}]
from radical.entk import Task, Stage
s = Stage()   
t = Task()
<add task properties>
s.add_tasks(t)
s.post_exec = {
                'condition': <function_1 name>,
                'on_true':   <function_2 name>,
                'on_false':  <function_3 name>
              }
\end{lstlisting}
\end{figure}

Users define the three functions specified as post-execution properties of a 
Stage or Pipeline, based on the requirements of their application. As such,
these functions can modify the existing TG or extend it as per the three
adaptivity types described in \S\ref{ssec:adap_types}.

Ref.~\cite{van2000dealing} specifies multiple strategies to perform
adaptation: forward recovery, backward recovery, proceed, and transfer. In
\entk, we implement a non-aggressive adaptation strategy, similar to
`transfer', where a new TG is created by modifying the current TG only after
the completion of part of that TG\@. The choice of this strategy is based on
the current science drivers where tasks that have already executed and tasks
that are currently executing are not required to be adapted but all
forthcoming tasks might be.

Modifying the TG at runtime requires coordination among EnTK components to
ensure consistency in the TG representation. AppManager holds the global view of
the TG and, upon instantiation, Workflow Processor maintains a local copy of
that TG\@. The dequeue sub-component of Workflow Processor acquires a lock
over the local copy of the TG, and invokes the adaptation operation as
described by the post-execution property of stages and pipelines. If the local
copy of the TG is modified, Workflow Processor transmits those changes to
AppManager that modifies the global copy of TG\@, and releases the lock upon
receiving an acknowledgment. This ensures that adaptations to the
TG are consistent across all components, while requiring minimal communication.

Pipeline, stage, and task descriptions alongside the specification of an
adaptation operation as post-execution for pipelines and stages enable the
expression of adaptive workflows. The `transfer' strategy enacts the
adaptivity of the TG, and the implementation in \entk ensures consistency and
minimal communication in executing adaptive workflows. Note how the design
and implementation of adaptivity in \entk does not depend on specific
capabilities of the software package executed by each task of the ensemble
workflow.

\section{Experiments}\label{sec:exps}
We perform three sets of experiments. The first set characterizes the
overhead of \entk{} when performing the three types of adaptation described
in \S\ref{ssec:adap_types}. The second set validates our implementation of
the two science drivers presented in \S\ref{sec:science_motiv} against
reference data. The third set compares our implementation of adaptive expanded
ensemble algorithm with local and global analysis against results obtained with 
a single and an ensemble of MD simulations.

We use four application kernels in our experiments:
\texttt{stress-ng}~\cite{stressng},
\texttt{GROMACS}~\cite{abraham2015gromacs}, \texttt{OpenMM}~\cite{openmm} and 
Python scripts. \texttt{stress-ng} allows to control the computational duration
of a task for the experiments that characterize the adaptation overhead of 
\entk{}, while \texttt{GROMACS} and \texttt{OpenMM} are the simulation kernels
for the expanded ensemble and Markov state modeling validation experiments.

We executed all experiments from the same host machine but we targeted three
HPC systems, depending on the amount and availability of the resources
required by the experiments, and the constraints imposed by the queue policy
of each machine. NCSA Blue Waters and ORNL Titan were used for characterizing
the adaptation overhead of \entk{}, while XSEDE SuperMIC was used for the
validation and production scale experiments.

\subsection{Characterization of Adaptation Overhead}

\begin{table*}
    \caption{\footnotesize Parameters of the experiments plotted in
    Fig.~\ref{fig:overheads}}\label{tab:experiments}
    \centering
    \begin{tabular}{l 
                    l 
                    l 
                    l 
                    l 
                    }
        \toprule
        \B{ID}                                 &
        \B{Figure}                             &
        \B{Adaptation Type}                    & 
        \B{Experiment variable}                & 
        \B{Fixed parameters}                   \\
        \midrule
        \B{I}                                  &  
        \ref{fig:overheads}i                   &  
        Task-count                             &  
        Number of adaptations                  &  
        \makecell[cl]{
        Number of tasks added per adaptation = 16, \\
        Type of tasks added = single-node}     \\ 
        \midrule
        \B{II}                                 &  
        \ref{fig:overheads}ii                  &  
        Task-count                             &  
        \makecell[cl]{
        Number of tasks added \\
        per adaptation}                        &  
        \makecell[cl]{
        Number of adaptations = 2, \\
        Type of tasks added = single-node}     \\ 
        \midrule
        \B{III}                                &  
        \ref{fig:overheads}iii                 &  
        Task-count                             &  
        Type of tasks added                    &  
        \makecell[cl]{
        Number of adaptations = 2, \\
        Number of tasks added per 
        adaptation = $2^{10}*2^{s}$ 
        (s=stage index)}                       \\ 
        \midrule
        \B{IV}                                 &  
        \ref{fig:overheads}iv                  &  
        Task-order                             &  
        Number of adaptations                  &  
        \makecell[cl]{
        Number of re-ordering operations 
        per adaptation = 1, \\
        Type of re-ordering = uniform shuffle} \\ 
        \midrule
        \B{V}                                  &  
        \ref{fig:overheads}v                   &  
        Task-property                          &  
        Number of adaptations                  &  
        \makecell[cl]{
        Number of property modified per
        adaptation = 1, \\ 
        Property adapted = Number of
        cores used per task}                   \\ 
        \bottomrule
    \end{tabular}
\end{table*}

We perform five experiments to characterize the overhead of adapting ensemble
workflows encoded using \entk{}. Each experiment measures the overhead of a
type of adaptation as a function of the number of adaptations. In the case of
task-count adaptation, the overhead is measured also as a function of the
number of tasks and of their type, single- or multi-node. This is relevant
because with the growing of the size of the simulated molecular system and of
the duration of that simulation, multi-node tasks may perform better than
single-node ones.

Each experiment measures \entk{} Adaptation Overhead and Task Execution Time.
The former is the time taken by \entk{} to adapt the workflow by invoking
user-specified algorithms; the latter is the time taken to run the executables
of all tasks of the workflow. Consistent with the scope of this paper, the
comparison between each adaptation overhead and task execution time offers a
measure of the efficiency with which \entk{} implements adaptive
functionalities. Ref.~\cite{power-of-many17} has a detailed analysis of other
overheads of \entk{}.

Table~\ref{tab:experiments} describes the variables and fixed parameters of
the five experiments about adaptivity overheads in \entk{}. In these
experiments, the algorithm is encoded in \entk{} as 1 pipeline consisting of
several stages with a set of tasks. In the experiments I--III about
task-count adaptation, the pipeline initially consists of a single stage with
16 tasks of a certain type. Each adaptation, at the completion of
a stage, adds 1 stage with a certain number of tasks of a certain type,
thereby increasing the task-count in the workflow.

In experiments IV--V, the workflow is encoded as 1 pipeline with 17, 65,
or 257 stages with 16 tasks per stage. Each adaptation occurs upon the
completion of a stage and, in the case of task-order adaption, the remaining
stages of a pipeline are shuffled. In the case of task-property adaption,
the number of cores used by the tasks of the next stage is set to a random
value below 16, keeping the task type to single-node. The last stage of both
experiments are non-adaptive, resulting in 16, 64, and 256 total adaptations.

In the experiments I, IV and V, where the number of adaptations varies, each
task of the workflow executes the \texttt{stress-ng} kernel for 60 seconds.
For the experiments II and III with \(O(1000)\) tasks, the execution duration
is set to 600 seconds so to avoid performance bottlenecks in the underlying
runtime system and therefore interferences with the measurement of \entk{}
adaptation overheads. All experiments have no data movement as the
performance of data operations is independent from that of adaptation.

\begin{figure*}
	\includegraphics[width=\textwidth]{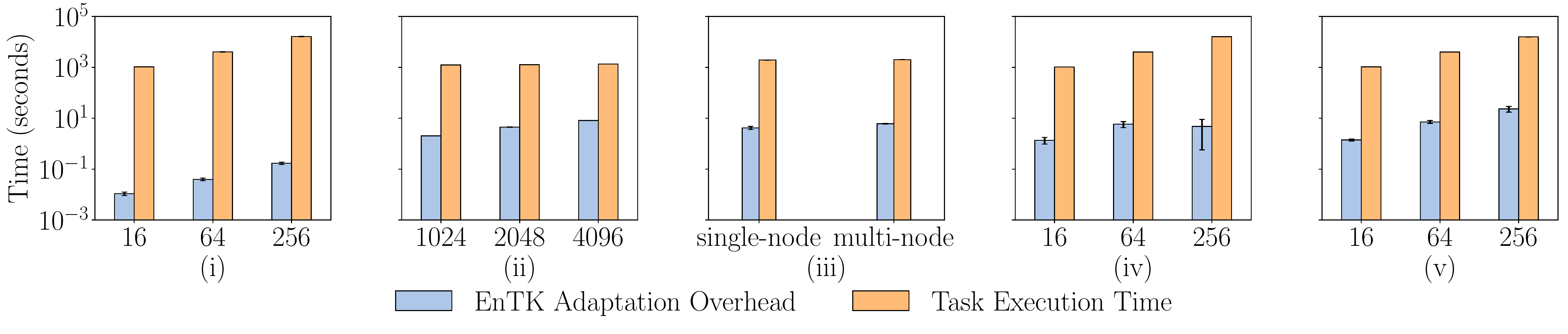}
	\Description{Five bar plots, each with number of nodes on the x-axes and
	time in seconds on the y-axes. Two bars: \entk{} Adaptation Overhead and
	Task Execution Time. In all plots, the former is shorter than the
	latter.}
	\caption{\footnotesize \entk{} Adaptation Overhead and Task Execution
	Time for task-count (i, ii, and iii), task-order (iv), and task-property
	(v) adaptations.}\label{fig:overheads}
\end{figure*}

Figs.~\ref{fig:overheads}(i),~\ref{fig:overheads}(iv), and~\ref{fig:overheads}(v)
show that \entk{} Adaptation Overhead and Task Execution Time increase
linearly with the increasing of the number of adaptations. \entk{} Adaptation
Overhead increases due to the time taken to compute the additional
adaptations and its linearity indicates that the computing time of each
adaptation is constant. Task Execution Time increases due to the time taken
to execute the tasks of the stages that are added to the workflow as a result
of the adaptation.

Figs.~\ref{fig:overheads}(i),~\ref{fig:overheads}(iv), and~\ref{fig:overheads}(v)
also show that task-property adaptation (v) is the most expensive, followed
by task-order adaptation (iv) and task-count (i) adaptation. These
differences depend on the computational cost of the Python functions executed
during adaptation: in task-property adaptation, the function parses the
entire workflow and invokes the Python \texttt{random.randint} function 16
times per adaptation; in task-order adaptation, the Python function shuffles
a Python list of stages; and in task-count adaption, the Python function
creates an additional stage, appending it to a list.

In Fig.~\ref{fig:overheads}(ii), \entk{} Adaptation Overhead increases linearly
with an increase in the number of tasks added per task-count adaptation,
explained by the cost of creating additional tasks and adding them to the
workflow. The Task Execution Time remains constant at \(\approx 1200s\),
since sufficient resources are acquired to execute all the tasks
concurrently.

Fig~\ref{fig:overheads}(iii) compares \entk{} Adaptation Overhead and Task
Execution Time when adding single-node and multi-node tasks to the workflow.
The former is greater by \(\approx 1s\) when adding multi\-node tasks,
whereas the latter remains constant at \(\approx 1200s\) in both scenarios.
The difference in the overhead, although negligible when compared to Task
Execution Time, is explained by the increased size of a multi-node task
description. As in Fig.~\ref{fig:overheads}(ii), Task Execution Time remains
constant due to availability of sufficient resources to execute all tasks
concurrently.

Experiments I--V show that \entk{} Adaptation Overhead is proportional to the
computing required by the adaptation algorithm and is not determined by the
design or implementation of \entk{}. In absolute terms, \entk{} Adaptation
Overhead is orders of magnitude smaller than Task Execution Time. Thus,
\entk{} advances the practical use of adaptive ensemble workflows.

\subsection{Validation of Science Driver Implementations}

We implement the two science drivers of \S\ref{sec:science_motiv} using the
abstractions developed in \entk{}. We validate our implementation of Expanded
Ensemble (EE) by calculating the binding of the cucurbit[7]uril
6-ammonio-1-hexanol host-guest system, and our implementation of Markov State
Modeling (MSM) by simulating the Alanine dipeptide system and comparing our
results with the reference data of the DESRES group~\cite{msm-ref-data}.

\subsubsection{Expanded Ensemble}

We execute the EE science driver described in \S\ref{ssec:usecase_ee} on
XSEDE SuperMIC for a total of 2270ns MD simulation time. To validate the
process, we carry out a set of simulations of the binding of cucurbit[7]uril
(host) to 6-amino-1-hexanol (guest) in explicit solvent for a total of
29.12ns per ensemble member, and compare the final free energy estimate to a
reference calculation. Each ensemble member is encoded in \entk{} as a
pipeline of stages of simulation and analysis tasks, where each pipeline uses
1 node for 72 hours. With 16 ensemble members (i.e., pipelines) for the
current physical system, we use \(\approx 1K/23K\) node/core-hours of
computational resources.

The EE simulates the degree of coupling between the guest and the rest of the
system (water and host). As the system explores the coupling using EE
dynamics, it binds and unbinds the guest to and from the host. The free
energy of this process is gradually estimated over the course of the
simulation, using the Wang-Landau
algorithm~\cite{wang-landau:prl:2001:wang-landau}. However, we hypothesize
that we can speed convergence by allowing parallel simulations to share
information with each other, and estimate free energies using the potential
energy differences among states and the Multistate Bennett Acceptance Ratio
(MBAR) algorithm~\cite{shirts-chodera:jcp:2008:mbar}.

We consider four variants of the EE method:
\begin{compactitem}
	\item \textbf{Method 1:} one continuous simulation, omitting \emph{any}
	intermediate analysis.
	\item \textbf{Method 2:} multiple parallel simulations without \emph{any}
	intermediate analysis.
	\item \textbf{Method 3:} multiple parallel simulations with local
	intermediate analysis, i.e., using current and historical simulation
	information from only its own ensemble member.
	\item \textbf{Method 4:} multiple parallel simulations with global
	intermediate analysis, i.e., using current and historical simulation
	information from all ensemble members.
\end{compactitem}

In each method, the latter 2/3 of the simulation data available at the time
of each analysis is used for free energy estimates via the MBAR algorithm. In
methods 3 and 4, adverse effects of the Wang-Landau algorithm are eliminated
due to the intermediate analyses. These provide a better estimate of the
weights that are used to force simulations to visit desired distributions in the
simulation condition space (see \S\ref{ssec:usecase_ee}). Note that in methods
3 and 4, where intermediate analysis is used to update the weights, the
intermediate analysis is always applied at 320ps intervals.


The reference calculation consisted of four parallel simulations that ran for
200ns each and with fixed weights, i.e., using a set of estimated weights and
not using the Wang-Landau algorithm. MBAR was used to estimate the free
energy for each of these simulations. 

\begin{figure}
	\centering
	\includegraphics[width=0.48\textwidth]{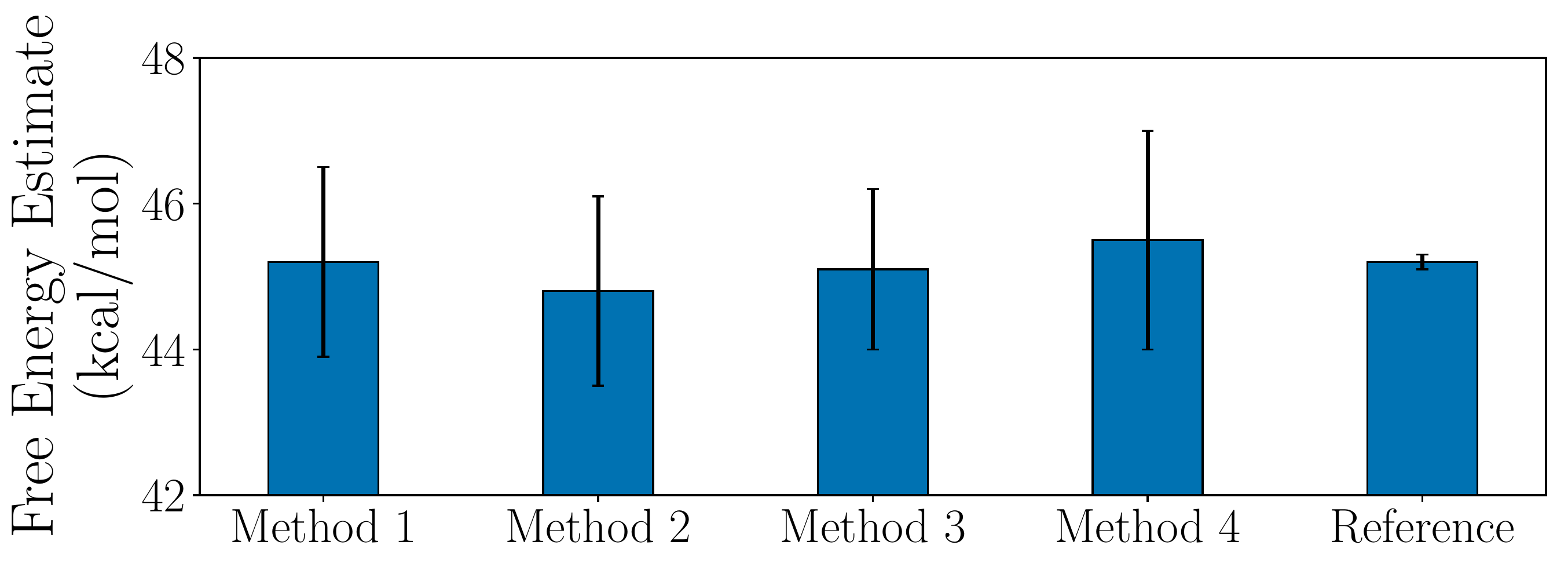}
	\Description{One bar plot with 4 methods and 1 Reference categories in
	the x-axes and free energy estimate (Kcal/mol) on the y-axes. Every bar
	has an error bar.}
	\caption{\footnotesize Validation of EE implementation: Observed
	variation of free energy estimate for methods 1--4. Reference is the MBAR
	estimate and standard deviation of four 200ns fixed weight
	expanded-ensemble simulations.}\label{fig:ee-validation}
\end{figure}

Fig.~\ref{fig:ee-validation} shows the free energy estimates obtained through
each of the four methods with the reference calculation value. Final
estimates of each method agree within error to the reference value. Validating 
that the four methods used to implement adaptive ensembles converge the free
energy estimate to the actual value\@.

\subsubsection{Markov State Modeling}
We execute the MSM science driver described in
\S\ref{ssec:usecase_msm} on XSEDE SuperMIC for a total of
100ns MD simulation time over multiple iterations. Each iteration of the TG is
encoded in \entk{} as one pipeline with 2 stages consisting of 10 simulation
tasks and 1 analysis task. Each task uses 1 node to simulate 1ns.

We compare the results obtained from execution of the \entk{} implementation
against reference data by performing the clustering of the reference data and
deriving the mean eigenvalues of two levels of the metastable states, i.e.,
macro- and micro-states. The reference data was generated by a non-adaptive
workflow consisting of 10 tasks, each simulating 10ns.

\begin{figure}
 \centering
 \includegraphics[width=0.48\textwidth]{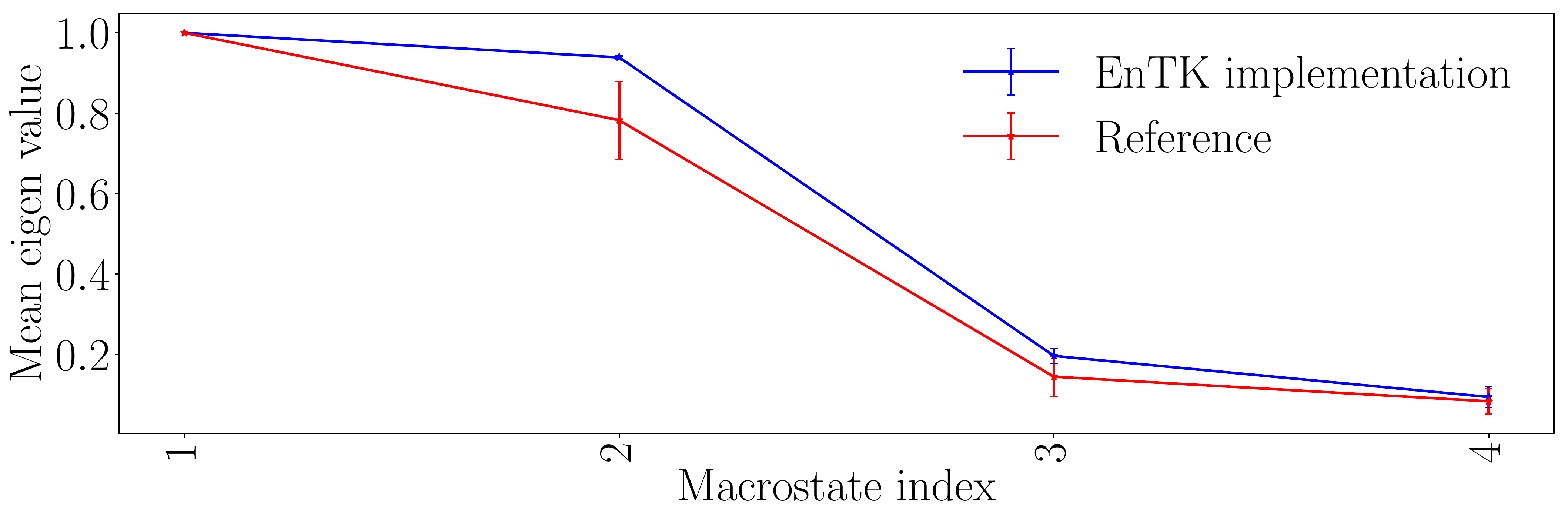}
 \includegraphics[width=0.48\textwidth]{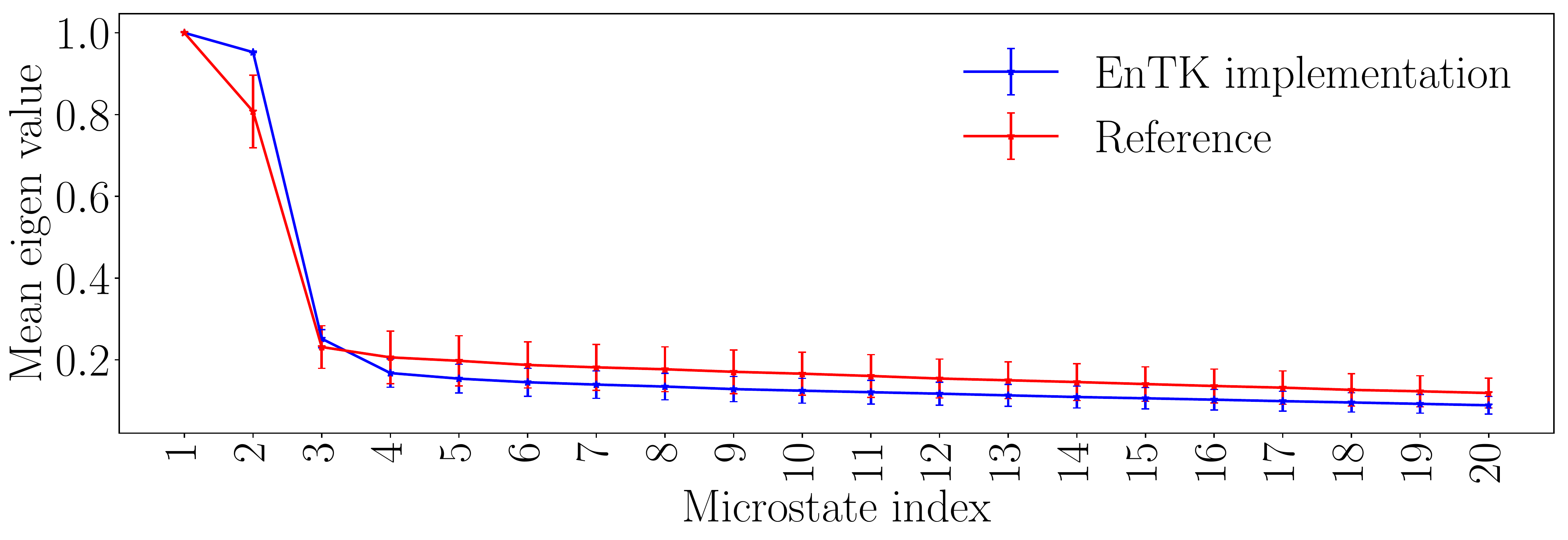}
 \Description{Two line plots. The top one has Macrostate Index on the x-axes,
 the bottom one Microstate Index; both have Mean Eigen Value on the y-axes.
 Two lines per plot representing \entk{} implementation and Reference.}
 \caption{\footnotesize Mean eigenvalue attained by the macro-states (top)
 and micro-states (bottom) by Alanine dipeptide after aggregate simulation
 duration of 100ns implemented using \entk{} compared against reference
 data.}\label{fig:msm-validation}
\end{figure}

Eigenvalues attained by the macro-states (top) and micro-states (bottom) in
the \entk{} implementation and reference data are plotted as a function of the
state index in Fig.~\ref{fig:msm-validation}. Final eigenvalues attained by
the implementation agree with the reference data within the error bounds. The
validation of the implementation warrants that similar implementations should
be investigated for larger molecular systems and longer durations, where the
aggregate duration is unknown and termination conditions are evaluated during
runtime.

\subsection{Evaluation of Methodological Efficiency using Adaptive Capabilities in \entk}

\begin{figure}
	\centering
	\includegraphics[width=0.48\textwidth]{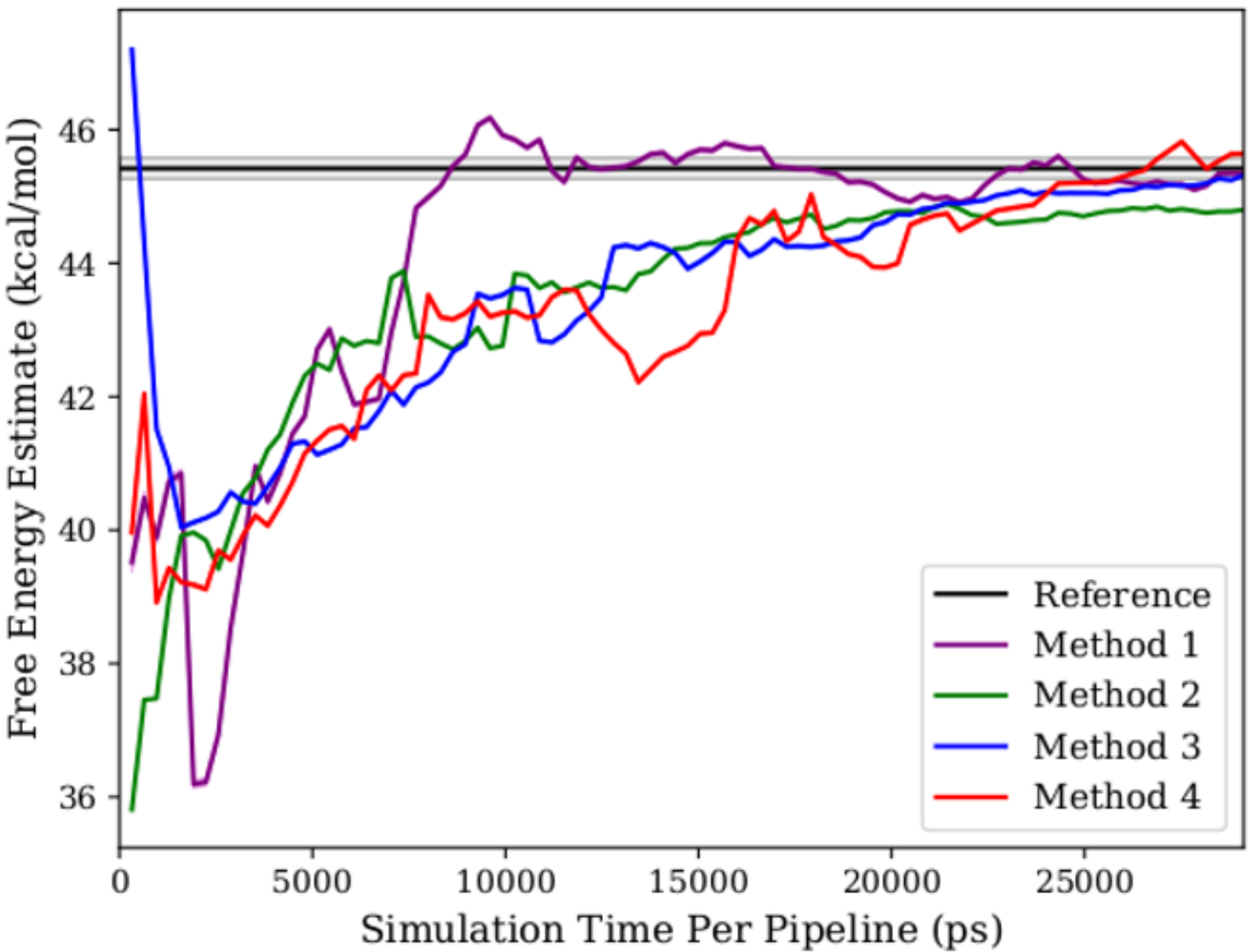}
	\Description{Line plot with Simulation Time per Pipelines on the x-axes
	and Free Energy Estimate (Kcal/mol) on the y-axes.5 lines: Reference and
	Method 1 to 4.}
	\caption{\footnotesize Convergence of expanded ensemble 
	implementation: Observed convergence behavior in methods 1--4. Reference
	is the MBAR estimate of the pooled data and the standard deviation of the
	non-pooled MBAR estimates of four 200ns fixed weight expanded ensemble
	simulations.}\label{fig:ee-convergence}
   \end{figure}

We analyzed the convergence properties of the free energy estimate using the
data generated for the validation of EE\@. 
%
%
The convergence behavior of Method 1 observed in
Fig.~\ref{fig:ee-convergence} implies that the current method converges
faster than ensemble based methods but does not represent the average
behavior of the non-ensemble based approach. The average behavior is depicted
more clearly by Method 2 because this method averages the free energy
estimate of 16 independent single simulations.

The most significant feature of Fig.~\ref{fig:ee-convergence} is that all
three ensemble based methods converge at similar rates to the reference
value. We initially hypothesized that adding adaptive analysis to the
estimate of the weights would improve convergence behavior but we see no
significant change in these experiments. However, the methodology described
here gives researchers the ability to implement additional adaptive elements
and test their effects on system properties. Additionally, these adaptive
elements can be implemented on relatively short time scales, giving the
ability to test many implementations.

Analysis of these simulations revealed a fundamental physical reason that
demonstrates a need for additional adaptivity to successfully accelerate
these simulations.  Although expanded ensemble simulations allowed the ligand
to move in and out of the binding pocket rapidly, the slowest motion,
occurring on the order of 10s of nanoseconds, was the movement of water out
of the binding pocket, allowing the ligand to bind as water backs into a
vacant binding pocket. Simulation biases that equilibrate on shorter
timescales may stabilize either the waters out or the waters in
configurations, preventing the sampling of both configurations.
Additional biasing variables are needed to algorithmically accelerate this
slow motions, requiring a combination of metadynamics and expanded ensemble
simulations, with biases both in the protein interaction variable and the
collective variable of water occupancy in the binding pocket. Changes in the
PLUMED2 metadynamics code are being coordinated with the
developers to make this possible.

Analysis of the slow motions of the system suggests the potential power of
more complex and general adaptive patterns. Simulations with accelerated
dynamics along the hypothesized degrees of freedom can be carried out, and
resulting dynamics can be analyzed, automated and monitored for degrees of
freedom associated with remaining slow degrees of motion~\cite{tiwary_2016}.
Accelerated dynamics can be adaptively adjusted as the simulation process
continues. Characterization experiments suggest that \entk{} can support the
execution of this enhanced adaptive workflow with minimal overhead.

\section{Conclusion}
Scientific problems across domains such as biomolecular science, climate
science and uncertainty quantification require ensembles of computational
tasks to achieve a desired solution. Novel approaches focus on adaptive
algorithms that leverage intermediate data to study larger problems, longer
time scales and to engineer better fidelity in the modeling of complex
phenomena. In this paper, we described the operations in executing adaptive
workflows, classified the different types of adaptations, and described
challenges in implementing them in software tools. We enhanced EnTK to
support the execution of adaptive workflows on HPC systems. We characterized
the adaptation overhead in EnTK, validated the implementation of the two
science drivers and executed expanded ensemble at production scale,
evaluating its sampling capabilities. To the best of our knowledge, this is
the first attempt at describing and implementing multiple adaptive ensemble
workflows using a common conceptual and implementation framework.


\printbibliography

\end{document}